\documentclass[aps,prb,twocolumn,superscriptaddress]{revtex4-1}

\usepackage{epsfig}
\usepackage{amssymb} 
\usepackage{amsfonts} 
\usepackage{mathtools} 
\usepackage{dcolumn} 
\usepackage{graphicx} 
\usepackage{color}  
\usepackage{bm}  
\usepackage{comment} 
\usepackage{units}
\usepackage{sidecap} 
\usepackage{textcomp} 
\usepackage{units}
\usepackage{sidecap} 
\usepackage{textcomp} 
\usepackage{mathrsfs} 

\usepackage{array} 
\usepackage{upgreek}
\usepackage{tabularx}
\usepackage{latexsym}
\usepackage{bm,array}
\usepackage{amsfonts}
\usepackage{amssymb}
\usepackage{amsmath}
\usepackage{bigstrut}
\usepackage{placeins}

\newcommand{\bpm}{\begin{pmatrix}}
\newcommand{\epm}{\end{pmatrix}}
\newcommand{\be}{\begin{eqnarray}}
\newcommand{\ee}{\end{eqnarray}}
\newcommand{\ba}{\begin{array}}
\newcommand{\ea}{\end{array}}

\def \lio{Li$_2$IrO$_3$}
\def \nio{Na$_2$IrO$_3$}

\def \alio{$\alpha$-Li$_2$IrO$_3$}
\def \blio{$\beta$-Li$_2$IrO$_3$}
\def \glio{$\gamma$-Li$_2$IrO$_3$}

\begin{document}
\title{Hidden spin-orbital order in the Kitaev hyperhoneycomb \blio}

\author{Alejandro Ruiz}
\affiliation{Department of Physics, University of California, Berkeley, California 94720, USA}
\affiliation{Materials Sciences Division, Lawrence Berkeley National Laboratory, Berkeley, California 94720, USA}

\author{Vikram Nagarajan}
\affiliation{Department of Physics, University of California, Berkeley, California 94720, USA}
\affiliation{Materials Sciences Division, Lawrence Berkeley National Laboratory, Berkeley, California 94720, USA}

\author{Mayia Vranas}
\affiliation{Department of Physics, University of California, Berkeley, California 94720, USA}
\affiliation{Materials Sciences Division, Lawrence Berkeley National Laboratory, Berkeley, California 94720, USA}

\author{Gilbert Lopez}
\affiliation{Department of Physics, University of California, Berkeley, California 94720, USA}
\affiliation{Materials Sciences Division, Lawrence Berkeley National Laboratory, Berkeley, California 94720, USA}

\author{Gregory T. McCandless}
\affiliation{Department of Chemistry and Biochemistry, The University of Texas at Dallas, Richardson, Texas 75080, USA}

\author{Itamar Kimchi}
\affiliation{JILA, NIST and Department of Physics, University of Colorado, Boulder, Colorado 80309, USA}

\author{Julia Y. Chan}
\affiliation{Department of Chemistry, The University of Texas at Dallas, Richardson, Texas 75080, USA}

\author{Nicholas P. Breznay}
\affiliation{Department of Physics, Harvey Mudd College, Claremont, California 91711, USA}

\author{Alex Fra\~{n}\'o}
\affiliation{Department of Physics, University of California, San Diego, California 92093, USA}

\author{Benjamin A. Frandsen }
\affiliation{ Department of Physics and Astronomy, Brigham Young University, Provo, Utah 84602, USA}

\author{James G. Analytis}
\affiliation{Department of Physics, University of California, Berkeley, California 94720, USA}
\affiliation{Materials Sciences Division, Lawrence Berkeley National Laboratory, Berkeley, California 94720, USA}

\date{\today}

\begin{abstract}

We report the existence of a phase transition at high temperature in the 3D Kitaev candidate material, \blio. We show that the transition is bulk, intrinsic and orders a tiny magnetic moment with a spatially anisotropic saturation moment. We show that even though this transition is global, it does not freeze the local Ir moments, which order at much lower temperatures into an incommensurate state. Rather, the ordered moment has an orbital origin that is coupled to spin correlations, likely of a Kitaev origin. The separate ordering of spin-correlated orbital moments and of local Ir moments reveals a novel way in which magnetic frustration in Kitaev systems can lead to coexisting magnetic states.

\end{abstract}


\maketitle

\section{INTRODUCTION}

Since Khaliullin and Jackeli \cite{jackeli_mott_2009, chaloupka_kitaev-heisenberg_2010} first pointed out that Kitaev's frustrated compass model\cite{kitaev_anyons_2006} on a honeycomb lattice could be realized in 4d and 5d transition metal systems with octahedral co-ordination, such materials have become one of the most promising routes to experimentally realizing a quantum spin liquid. The ground state itself, first described by Kitaev \cite{kitaev_anyons_2006}, is characterized by the long range order of flux degrees of freedom, emerging from the fractionalization of the local spins into Majorana excitations. The ideal Kitaev model couples orthogonal directions of spin along the three different bond directions,
\begin{equation}
\mathcal{H} = K\sum_{\langle ij\rangle} S_i^\gamma S_j^\gamma
\end{equation}
where $\gamma = {x,y,z}$ specify the three compass directions of the Kitaev exchange, $K$. Importantly, in the \blio\,and \glio\,materials, one of these Kitaev axes is exactly parallel to the crystallographic $b$ axis.

Although some low-temperature signatures of novel excitations have been reported~\cite{kasahara_majorana_2018}, the magnetic order present in all candidate materials dominate most of their properties (zig-zag order in the case of $\alpha$-RuCl$_3$ and $\alpha$-\nio, and incommensurate order in $\alpha,\beta,\gamma-$\lio\,species)\cite{singh_antiferromagnetic_2010,PhysRevB.83.220403, PhysRevB.92.235119, biffin_noncoplanar_2014, biffin_unconventional_2014, williams_incommensurate_2016}. However, many recent studies have found high temperature signatures of these exotic states or proximity thereto. For example, recent spectroscopic and thermodynamic studies of $\alpha$-RuCl$_3$ \cite{sandilands_scattering_2015, banerjee_proximate_2016, banerjee_neutron_2017, do_majorana_2017, banerjee_excitations_2018, kasahara_unusual_thermal_hall_2018,PhysRevLett.114.147201, wang_direct_2018}, have reported evidence for the onset of nearest-neighbor Kitaev correlations, consistent with a proposal by Motome and co-authors of a thermal crossover from a paramagnet to a spin-``fractionalized" state \cite{nasu_vaporization_2014,nasu_thermal_frac_2015,nasu_fermionic_2016}. Similar studies have extended these conclusions to $\alpha,\beta,\gamma-$\lio~and $\alpha-$\nio~systems~\cite{mehlawat_HC_lio, glamazda_raman_2016,revelli_fingerprints_2019}.  

The nature of the ground state at these elevated temperatures is therefore of considerable interest~\cite{nasu_vaporization_2014,nasu_thermal_frac_2015,nasu_fermionic_2016,mehlawat_HC_lio, glamazda_raman_2016,revelli_fingerprints_2019}. However, due to the small size of the samples, relatively few studies have explored the three-dimensional \blio~materials in this temperature region. In this work, we focus on the magnetic and thermal properties of \blio, and reveal the presence of a phase transition at $\sim$100K. The transition is associated with the ferromagnetic-like ordering of a small moment, whose anisotropy closely follows the Kitaev principal axes. We argue that the properties of this state suggest the hidden order involves spin-correlated orbital moments, and not the local moments of the Ir ions. 




\begin{figure*}[ht]
    {\includegraphics[width=\textwidth]{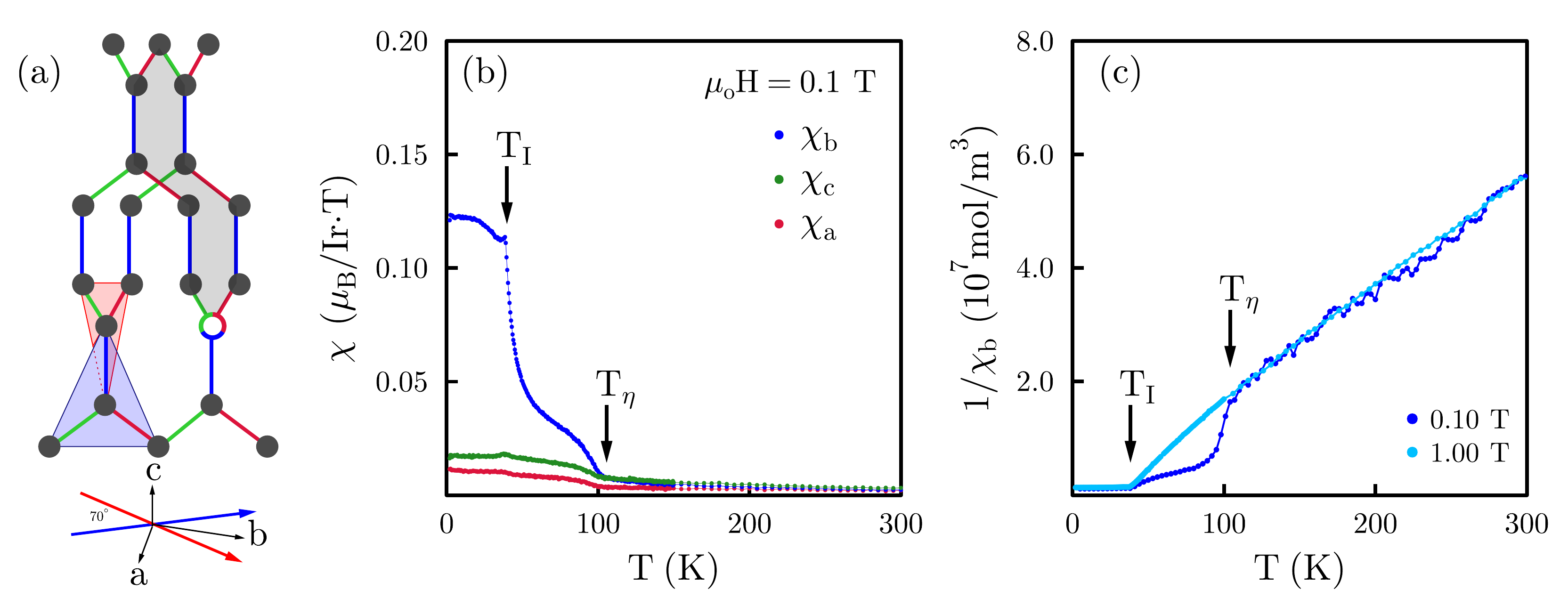}
  	\caption{(Color online) (a) Three dimensional structure of \blio, where the red, green and blue colors correspond to orthogonal compass directions of the Kitaev model.  The two triangles, situated 70$^\circ$ apart, show the possible  environments for a magnetic ion in \blio, and determine the $g$-factor anisotropy. Also shown is a site vacancy which can trap flux excitations in a Kitaev spin liquid, creating a large local moment. (b) The anisotropic magnetic susceptibility of \blio\ for an applied magnetic field of \unit[0.1]{T}. At $T_I=\unit[38]{K}$, the system transitions into an incommensurate spiral state with non-coplanar, counter-rotating moments. When a small magnetic field is applied ($H<\unit[0.5]{T}$), a separate transition is also observed at \unit[100]{K}. (c) Comparison of the inverse $\hat{b}$-axis susceptibility for  \unit[1.0]{T} and \unit[0.1]{T}. The low-field data shows two distinct behaviors: a linear response above \unit[100]{K} and a strong deviation from Curie-Weiss behavior $100 >T> \unit[40]{K}$.}
  	\label{fig:univ}} 
\end{figure*}

\section{EXPERIMENTAL RESULTS}

Single crystals of \blio\, were synthesized using standard techniques described in the SM (section S1). The 3D nature of \blio\ is realized in the hyperhoneycomb arrangement of the Ir atoms shown in Figure \ref{fig:univ}\,a, while the low-field anisotropic magnetic susceptibility is shown in Figure \ref{fig:univ}\,b. In Figure \ref{fig:univ}\,c, we contrast the inverse b-axis susceptibility measured at \unit[1]{T} and \unit[0.1]{T} to show that, above \unit[100]{K}, the magnetization is truly field independent  with an effective spin $\mathcal{J}=\nicefrac{1}{2}$, which can be completely understood as paramagnetic spins coupled to their orbital environment (see Supplementary Materials (SM) section S2A for details). Below $T_I=\unit[38]{K}$, the system orders into an incommensurate state with non-coplanar and counter-rotating spins ~\cite{biffin_unconventional_2014, biffin_noncoplanar_2014}. At $\sim\unit[100]{K}$ the principal axes of the magnetization re-order due to the presence of strong Kitaev-like correlations\cite{modic_lio}, such that the $b$-axis becomes dominant.  Our data shows that this re-ordering occurs due to the presence of a phase transition at $T_\eta$, which can only be observed using low applied magnetic fields. The smearing of this transition at higher fields (Fig. \ref{fig:univ}c) is likely why this transition has remained hidden in previous measurements (see also SM, section S2-S3)~\cite{modic_lio, takayama_hyperhoneycomb_2015,ruiz_correlated_2017, majumder_nmr_2019}. As the field decreases, this transition becomes apparent, as seen in the comparison data shown in Figure \ref{fig:univ}\,c. The magnetic signal is extremely reproducible between different samples and batches, and independent of the synthesis environment (crucible material or source of starting elements), and sample volume (see SM section S5). In addition, we find no evidence of competing crystalline phases in single crystal x-ray diffraction measurements (see SM, section S1 for details). Similar results were also observed in \glio, as described in SM, section S2. This suggests an impurity phase is extremely unlikely as an origin of this transition. Moreover, the transition temperature is conspicuously close to the temperature window under intense study in the 2D Kitaev candidate systems, where there is thought to be evidence of emergent, fractional excitations~\cite{mehlawat_HC_lio, glamazda_raman_2016,revelli_fingerprints_2019}. 

\begin{figure*}[ht]
    {\includegraphics[width=\textwidth]{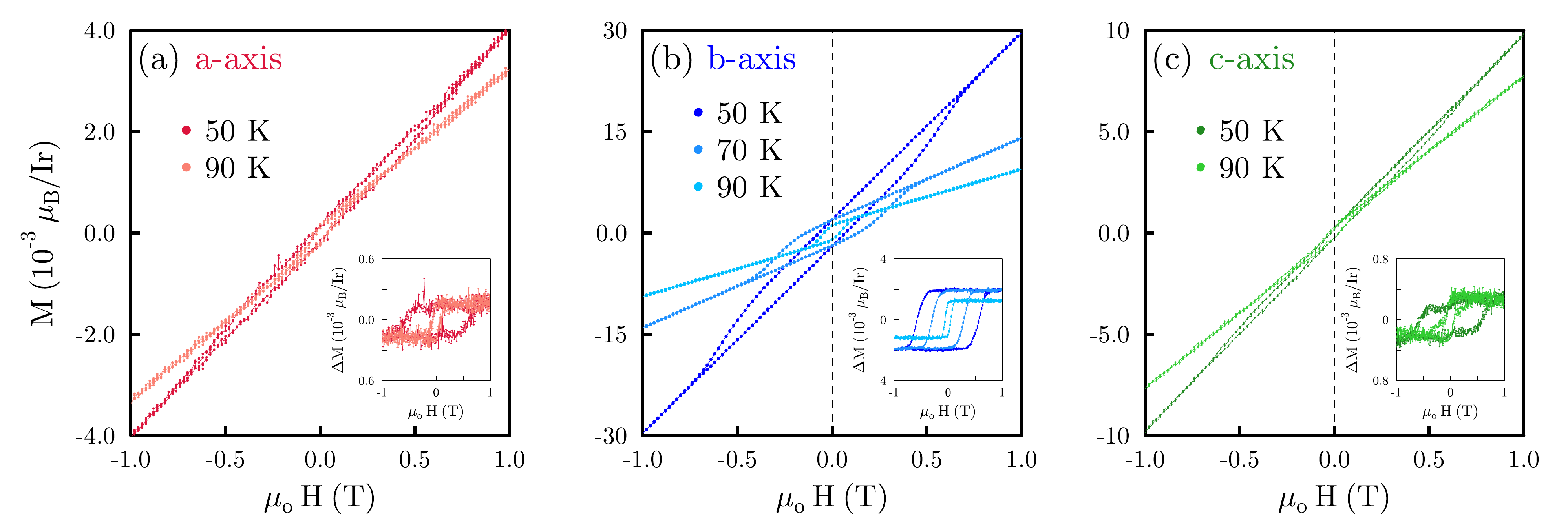}
  	\caption{(Color online) (a-c) Isotropy of the anisotropy field $H_a$ in $\beta$-Li$_2$IrO$_3$ along the three principal axes indicated. Hysteresis behavior was observed below $T_\eta=\unit[100]{K}$. The inset shows the data after the linear background corresponding to the high-field susceptibility is subtracted. }
  	\label{fig:coerc}}
\end{figure*}

\begin{figure*}[ht]
    {\includegraphics[width=\textwidth]{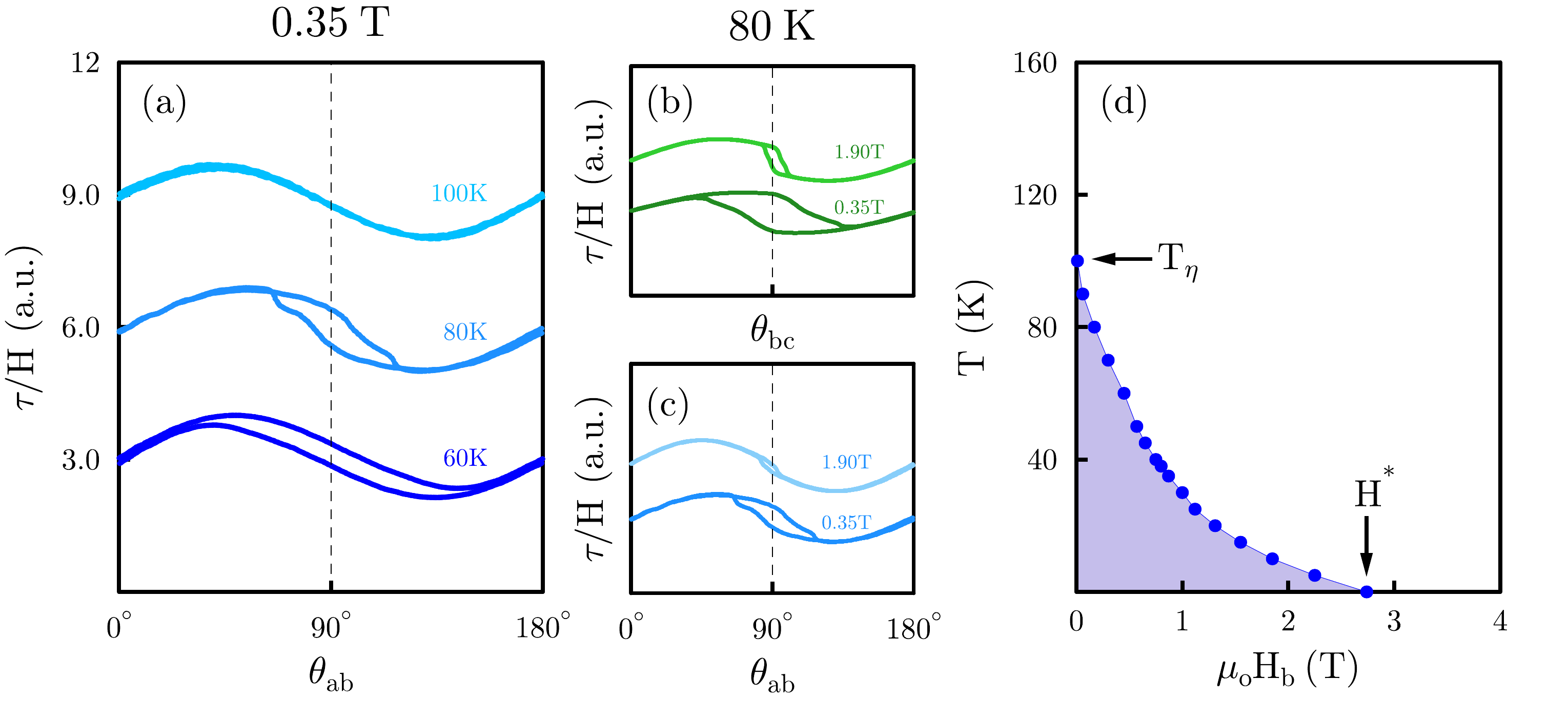}
    \caption{(Color online) (a) The angular dependent torque, $\tau_a=1/2(\chi_b-\chi_c)H^2\sin{2\theta_{ab}}$, also shows hysteretic behavior below $T_\eta$ for an applied field $H=\unit[0.35]{T}$  (b,c) Hysteresis is observed for rotations in the $ab$ and $bc$ planes, but not in the $ac$ plane. This behavior disappears at high fields, as it is evident in the data presented for $\unit[80]{K}$. In this case $\theta=0$ corresponds to $H\parallel b$ for rotations in the $bc$ and $ab$ planes. (d) The anisotropy field $H_a$ was extracted from M(H) and $\tau/H$($\theta$) measurements. $H_a$ appears to be indifferent to the low temperature phase boundary, and it terminates $H_a(0)\sim H^*(0)$, suggesting the incommensurate and hidden phase might share a common energy scale. (d) The temperature dependence of the anisotropy field $H_a$, see SM section S4 for low temperature determination.}
  	\label{fig:thermo}}
\end{figure*}

Figure\,\ref{fig:coerc}\,a-c shows the field-dependent magnetization below $T_\eta$ along three crystallographic directions, illustrating clear hysteresis behavior, and a coercive field that increases with decreasing temperature (in our case we parameterize this with the anisotropy field $H_a$, whose temperature dependence is shown in Figure \ref{fig:thermo}d). The insets in Figure \ref{fig:coerc} show the hysteresis curve after a linear background was subtracted, determined from the high field susceptibility in Figure \ref{fig:univ}\,c. 

The spatial dependence of the anisotropy field $H_a$ is independent of crystallographic direction, which is very surprising given the anisotropic nature of the crystal structure and magnetism. In contrast, the saturation moment $M_s$ appears to vary by a factor of $\sim10$, mirroring the anisotropy of the susceptibility, which is thought to originate from the presence of Kitaev correlations~\cite{kimchi_unified_2015,modic_lio}. We note that while this background subtraction makes the precise determination of the saturation moment difficult, the hysteresis loops in any direction rise with approximately the same gradient, suggesting they approach saturation with the same functional form. This implies $M_s$ must be strongly spatially anisotropic. In typical magnetically ordered systems, or even in spin glasses, the behavior is usually the other way around, where the saturation moment is isotropic (since it is related to the local moment), while the coercive field is anisotropic (since it is related to the anisotropy of the free energy and/or structural anisotropies of domain boundaries) \cite{blundell_magnetism_2001}. The spatial anisotropy of $M_s$ suggests a strong orbital component to the magnetic species that orders at $T_\eta$.

 Figure \ref{fig:thermo} shows the angular dependence of the magnetic torque of \blio\ at fixed fields and temperatures, respectively. Figure\,\ref{fig:thermo}\,a displays the onset of hysteretic behavior in the $ab$ plane upon cooling below \unit[100]{K} when sweeping angle from \unit[0]{$^\circ$} to \unit[180]{$^\circ$} and back. Upon lowering temperature further, hysteresis occurs in a wider angular range, corresponding to the larger anisotropy field and the larger angle needed to allow for a greater component of $H$ along $b$. Hysteresis is observed for $H$ aligned in both the $bc$ and $ab$ planes, as seen in Figure\,\ref{fig:thermo}\,b,c. With increasing field, the angular range of hysteresis decreases until eventually it disappears. 

\begin{figure*}[ht]
    {\includegraphics[width=\textwidth]{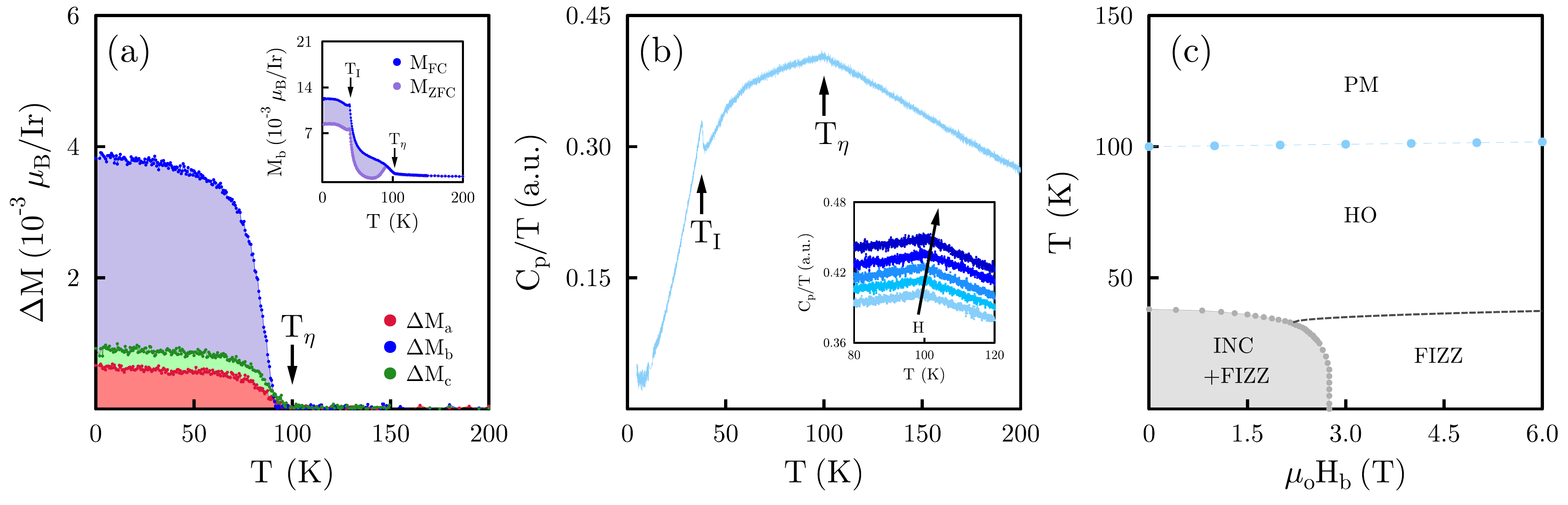}
  	\caption{(Color online) (a) Figure inset shows the field-cooled (FC) and zero field-cool (ZFC) magnetization with a $\unit[0.1]{T}$ field applied along the $b$-axis, while the main panel shows the magnetic irreversibility ,$\Delta M=M_{FC}-M_{ZFC}$. (b) Heat capacity response of \blio\, measured using the AC technique. A triangular cusp at $T_I=\unit[38]{K}$ shows the transition into the incommensurate spiral state while a small kink was observed at $T_\eta=\unit[100]{K}$. Inset shows temperature dependence of the heat capacity in fixed fields 0T, 2T, 4T, 6T and 8T in the direction of the arrow. $T_\eta$ appears to slightly increase with increasing field. (c) T-H phase diagram where the blue dots denote the hidden order (HO) as observed in the heat capacity, the gray dots represent H$^*$, the incommensurate state (INC) boundary, and the dotted line shows the field-induced zig-zag (FIZZ) state. For more information see reference \textcite{ruiz_correlated_2017}. }
  	\label{fig:HC}}
\end{figure*}

In Figure \ref{fig:HC}\,a, we show the zero field cooled (ZFC) and field cooled (FC) magnetization curves using \unit[0.1]{T} (inset), and their difference (main figure). The latter shows the natural form expected of a magnetic order parameter growing below $T_\eta$. Figure\,\ref{fig:HC}\,b shows the zero-field (AC) heat capacity response of \blio. Though absolute values of the specific heat are difficult to establish using AC techniques due to the frequency dependence of the response, the presence of a clear kink at  $T_\eta=\unit[100]{K}$ confirms the presence of a phase transition. DC heat capacity measurements, which yield more reliable absolute measures of the specific heat are on the other hand much less sensitive to weak phase transitions. Thus, although a sharp kink is not clearly visible, which is consistent with similar recent measurements~\cite{majumder_nmr_2019}, taking the difference between the heat capacity of \blio\,and its non-magnetic analogue $\beta-$Li$_2$PtO$_3$, shows that a measurable fraction of the magnetic degrees of freedom freeze out at $T_\eta$ (see SM Fig. S6). The phase transition at $T_\eta$ freezes very small fraction of the degrees of freedom, consistent with the smallness of the ordered moment itself.

The anisotropy field $H_a$, where the moment associated with the hidden phase is saturated, is shown by the blue dots in the phase diagram of Figure \,\ref{fig:thermo}\,d (see SM section S4 for low temperature determination of $H_a$). $H_a$ appears to be indifferent to the phase boundary as the system crosses into the incommensurate phase marked by $H^*$ (gray dots, Fig. \ref{fig:HC}c). On the other hand, $H_a$ terminates at the zero temperature at $H_a(0)\sim H^*(0)$ within experimental error, suggesting the incommensurate and hidden phase might share a common energy scale; the field required to polarize the hidden order is the same as that required to flip the incommensurate phase into the field induced zig-zag phase (FIZZ).

\begin{figure}[ht]
    {\includegraphics[width=\textwidth/2]{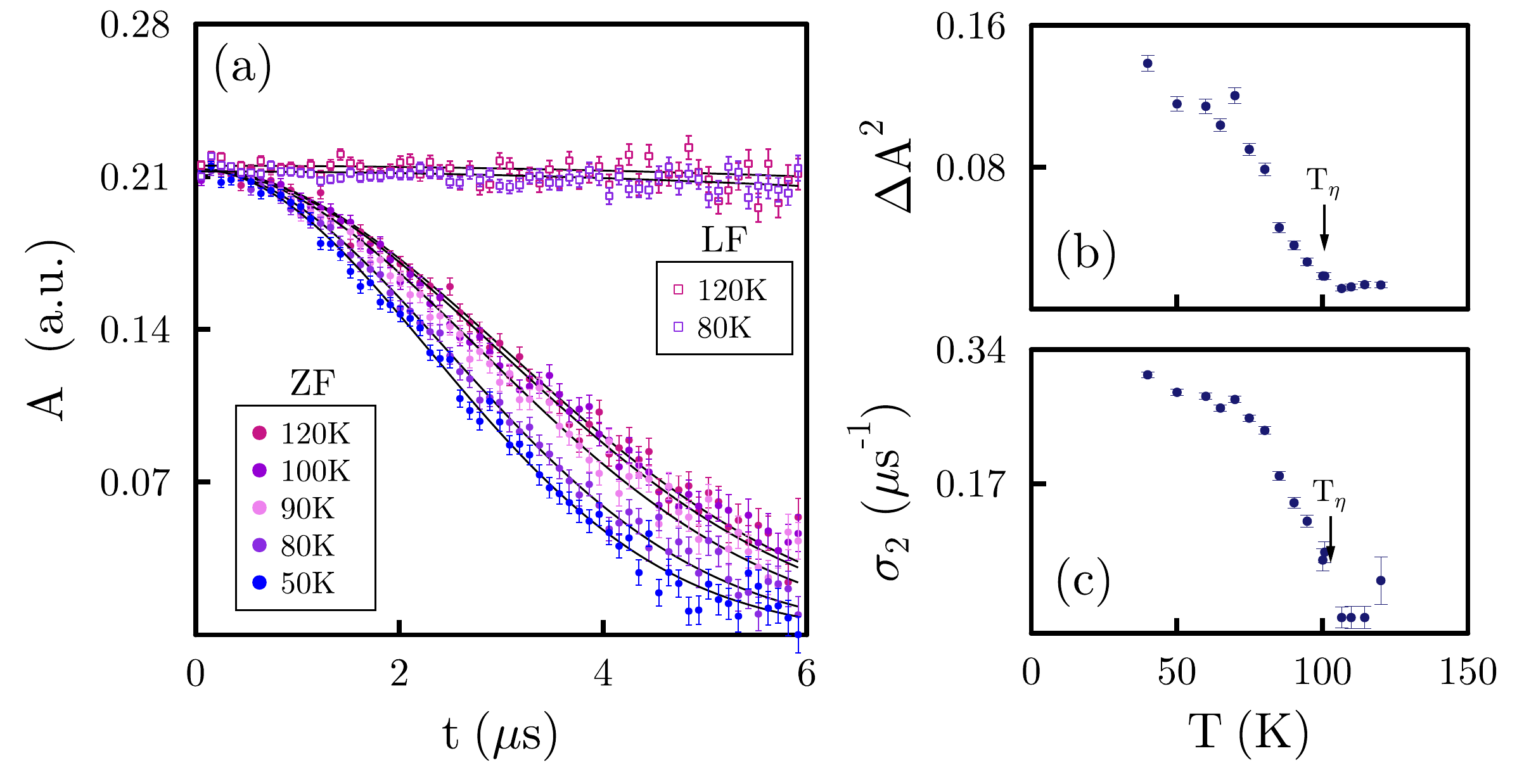}
  	\caption{(Color online) (a) $\mu$SR asymmetry, $A(t)$,  spectra at selected temperatures measured in zero field (ZF) and a \unit[50]{G} longitudinal field (LF). The ZF relaxation rate shows a clear increase below $T_\eta$. This relaxation can be decoupled with a very modest LF, indicating the development of weak, static magnetism. Solid lines are Gaussian fits to the data. (b) The asymmetry $A(t)=a_oe^{-t^2(\sigma_1^2+\sigma_2^2)/2}$ has two contributions: a T-independent nuclear contribution $\sigma_1$, and a T-dependent electronic contribution $\sigma_2$. The  T-dependent $\sigma_2$ evolves like an order parameter below $T_\eta$. A model independent metric, $\Delta A^2(T)=\sum_i{\frac{(A_i^{150K}-A_i^T)^2}{(A_i^{150K})^2}}$, is shown in panel (b), confirming the results of the fits. }
  	\label{fig:usr}}
\end{figure}

Our $\mu$SR measurements confirm that this feature in the magnetization and heat capacity data is intrinsic to \blio. As seen in Figure~\ref{fig:usr}\,a, our $\mu$SR results show a clear increase in the zero-field (ZF) relaxation rate at 100~K, precisely the same temperature at which the magnetization and heat capacity features were observed. A model-independent comparison of the asymmetry spectra reveals that the change in relaxation begins at 100 K and grows like an order parameter (Fig. ~\ref{fig:usr}\,b), which is confirmed by fitting a model and extracting the temperature-dependent relaxation rate (Fig. ~\ref{fig:usr}\,c). Details of the fits are given in the SM (section S6). The relaxation can be fully decoupled with a very modest longitudinal field of 50~G, indicating the development of weak, static magnetism in \blio\ below 100~K. We note that this type of magnetism is completely different from the long-range magnetically ordered state below $\sim$38~K in this system, which manifests in the $\mu$SR data as rapid oscillations and damping in the early-time portion of the asymmetry spectra \cite{choi_spin_2012,breakdown_pressure_2018,choi_fieldind_2019}. The $\mu$SR results are consistent with magnetization and heat capacity data which show the presence of a transition at $T_\eta$. 

The $\mu$SR data helps exclude an impurity origin of the hidden ordered phase (like the presence of inter-growths) since all or nearly all of the muons experience a change below $T_\eta$. Dynamics associated with Li-disorder as a possible origin for $T_\eta$ can also be ruled out; such disorder generally leads to exponential line-shapes in the asymmetry data, whereas the present data are clearly Gaussian \cite{ashton_muon_2014}. Finally, ordering of dilute magnetic impurities (which can lead to ferromagnetic transitions in magnetic semiconductors) can similarly be excluded since these lead to dramatic changes in the asymmetry data below the transition temperature, whereas we see a relatively small increase in the muon relaxation rate below $T_\eta$ (see Fig. \ref{fig:usr}a) \cite{dunsiger_spatially_2010,deng_liznmnas_2011,zhao_new_2013}. 


\section{DISCUSSION}

The thermodynamic and spectroscopic evidence unambiguously establishes the hidden order as an intrinsic thermodynamic phase in \blio; there exist sharp signatures in both susceptibility and heat capacity, and $\mu$SR shows the magnetic moment is static, existing throughout the volume of the sample. There are therefore two coexisting phases in this system: the incommensurate phase which onsets at $T_I=$\unit[38]{K}, and the hidden order at $T_\eta=$\unit[100]{K}. Strikingly, the hysteresis fields $H_a$ of the hidden phase crosses the incommensurate phase boundary in both field and temperature with complete impunity, suggesting they each order a distinct magnetic species. 

The existence of competing phases is widely known in these materials. In \blio\, for example, it is known that a zig-zag phase is close in energy and can be induced with the application of relatively small fields \cite{ruiz_correlated_2017, rousochatzakis_magnetic_2018,PhysRevB.97.125125}. However, the $\mu$SR data unambiguously rules this out, as the presence of such a phase would lead to oscillations in the muon relaxation. Another possibility is a valence-bond transition, similar to that seen under pressure in $\alpha$-RuCl$_3$ \cite{PhysRevB.97.241108} or \blio \cite{npb, veiga_pressure_2017, takayama_pressure_2019, veiga_dimerization_2019}. However the spin dimerization has an associated structural distortion that leads to strong hysteresis on warming and cooling, and this is absent in the current data. The $\mu$SR data is more consistent with a disordered magnet, like a spin glass. To explain our data, the moment of the disordered species would have to be extremely weak as, according to our fits, the local field is of the order of a few Gauss (by contrast the local field in \nio\,is an order of magnitude larger \cite{choi_spin_2012}). Even supposing that the true moment is somehow screened from the muons (which itself would require an exotic explanation given the absence of itinerant electrons to Kondo screen), the smallness of the induced moment in our magnetic measurements would suggest a highly dilute magnetic species. This, however, is difficult to reconcile with the high transition temperature, the sample-to-sample reproducibility, and the sharp heat capacity anomaly, all of which are rare in typical examples of dilute spin glasses \cite{Mydosh1993}. Moreover, the absence of relaxation effects, magnetic and thermal memory effects, and exchange bias is inconsistent with a spin glass scenario.

Moreover, the basic characteristics of the hidden order are inconsistent with a dilute magnetic semiconductor scenario, in which magnetic defects order ferromagnetically. The onset at 100K is much higher than the ordering of the large moment magnetic order appearing at $T_I=$38K. In particular, \blio\,is a Mott insulator with a local moment on every Ir site, as evident from 1/$T$ Curie-Weiss dependence and from the well studied spiral magnetic order, unlike a semiconductor. This is a crucial difference; magnetic dopant ions can be present here, and they can magnetize the local moment, but it seems highly unlikely that they give a give a ferromagnetic signature at temperatures much higher than the intrinsic large-moment magnetic order.

Nevertheless, there are other clues in the properties of the hidden phase that point to its origins. The saturation moment $M_{s}$, for example, is strongly anisotropic (Figure \ref{fig:coerc}). This conclusion can be seen to be independent of the background subtraction, since if $M_s$ was isotropic the hysteresis loops along each crystallographic direction would not be the same shape. Moreover, the isotropy of the hysteresis, parameterized by the field $H_{a}(i)$ ($i\in a,b,c$), illustrates that the hysteresis does not come from domain formation (which would be influenced by the orthorhombic structure), but from the anisotropy in the free energy itself. This can be seen by the following argument. In uniaxial ferromagnets, the anisotropy field is given by the ratio of the anisotropic free energy $K_a$ and the saturation moment $M_s$, so that the observation of an isotropic $H_{a} (i)\sim K_a(i)/M_{s}(i)$, suggests $M_{s}(i)$ follows the free energy anisotropy. From this we can make two conclusions. Firstly, the smallness of the $M_s$ and its spatial anisotropy strongly suggest an orbital origin. Secondly, this anisotropy exactly follows the magnetic principal axes and not the structural anisotropy of the orthorhombic crystal. Notably $M_s$ picks out the Kitaev $b$ axis as the dominant direction, just like the incommensurate phase. The magnetic species of the hidden order inherits signatures of Kitaev spin-spin correlations in the anisotropy of its energy landscape, but at the same time ordering a moment with a strongly orbital character, not of the local magnetic (Ir) ions.

Reconciling the dual character of the hidden order will require extensive future studies, but we speculate as to some possible scenarios here. For example, recent theoretical studies of Jahn-Teller distortions in related systems have shown the possible emergence of spin-nematic degrees of freedom. These could give rise to an emergent magnetic species that orders at relatively high temperatures \cite{Liu2019}, and couple together spin and orbital interactions, leading to nematic order with possibly the signatures we observe. However, we have performed structural refinements above and below $T_\eta$ and found no significant changes in the positions of any atomic species, suggesting weak Jahn-Teller effects (see SM S1). Another possibility is the scalar chiral spin order recently suggested as an explanation for the saw-tooth torque anomaly observed in RuCl$_3$ and \glio\cite{modic_chiral_2018}. We note that the anomalous torque onsets at exactly 100K, and extends into the incommensurate state. However, other studies have suggested that such anomalies can be understood by a field-dependent response with an anisotropic $g$-factor~\cite{riedl_saw-tooth_2018}. The association of a phase transition with the onset of the torque anomaly, reported here, should assist in distinguishing these scenarios.

Finally, we comment on an interesting possibility that might be a middle ground between these different scenarios. Recent theoretical studies of site-dilution in Kitaev spin liquids have revealed that vacancies form an emergent magnetic species\cite{Moessner2010b, Moessner2010} (Figure \ref{fig:univ}\,a). In this picture, the local fractionalization of spin degrees of freedom form moments in three dimensional systems that interact via the spin liquid \cite{Moessner2010b}. This may look like a disordered phase in a muon experiment, since there is no long range order of a local moment. However, such a phase could show a true thermal phase transition as the spin degrees of freedom fractionalize to form the medium through which these moments interact \cite{nasu_vaporization_2014}. We note that Raman spectroscopy in \blio\, has reported the presence of Fermionic excitations at finite energy (presumably arising from spin fractionalization), but not of a phase transition~\cite{glamazda_raman_2016}. On the other hand, evidence for such fractionalization in this temperature range has been reported in $\alpha$-RuCl$_3$ and \alio. \cite{sandilands_scattering_2015, banerjee_proximate_2016, banerjee_neutron_2017, do_majorana_2017, banerjee_excitations_2018, kasahara_unusual_thermal_hall_2018,PhysRevB.99.085136, PhysRevLett.120.217205,wang_direct_2018} The fact that this appears as a crossover in the $\alpha$-type structures and a phase transition in $\beta,\gamma$-type structures may simply reflect the different dimensionality of the materials.

\section{CONCLUSION}

In \blio, it is known that the principal axes of the magnetization undergo a dramatic reordering at $\sim$100K\cite{ruiz_correlated_2017, modic_lio}. Above this temperature they follow the structural anisotropy of the system, but as they approach $T_I$, they follow the spin-spin correlations of the system, which originate from a Kitaev term in the Hamiltonian~\cite{kimchi_three-dimensional_2014, kimchi_unified_2015} (see also an extended discussion in SM, section S2). Here we have shown that this reordering is actually accompanied by a bulk, intrinsic phase transition that is only visible at ultra-low fields, perhaps explaining why it has been hidden from previous measurements of this compound. The identity of this phase is unlikely to be one of the nearby ordered states known in the phase diagram of these systems, nor do its properties appear consistent with disordered phases like a typical spin glass. Rather, the observation of an anisotropic saturation moment that follows the Kitaev principal axes could arise if the ordered moment had an orbital origin that is tied to the spin-spin correlations of the Kitaev system. In this sense, the hidden order involves the ordering of a spin-correlated orbital magnetic species.
Given the intrinsic nature of the phase, we expect similar hidden states to appear in related materials which should be observable given sufficiently careful experiments in this temperature range.

\section{Acknowledgements}

The authors would like to thank Anthony Carrington, Chandra Varma, Natalia Perkins and Roser Valenti for fruitful discussions. In addition, material synthesis and experimental measurements were supported by the Department of Energy Early Career Award, Office of Basic Energy Sciences, Materials Sciences and Engineering Division, under Contract No. DE-AC02-05CH11231. A. Ruiz also acknowledges support from the National Science Foundation Graduate Research Fellowship under Grant No. DGE 1106400. M. Vranas acknowledges support from the UC LEADs program. N. P. Breznay was supported by the Gordon and Betty Moore Foundation’s EPiQS Initiative through Grant GBMF4374. Julia Y. Chan acknowledges NSF-DMR-1700030. I. Kimchi was supported by a National Research Council Fellowship through the National Institute of Standards and Technology.

\end{document}